\documentclass[%
 aip,
 amsmath,amssymb,
 reprint,%
]{revtex4-1}

\usepackage{graphicx}
\usepackage{dcolumn}
\usepackage{bm}

\usepackage[utf8]{inputenc}
\usepackage[T1]{fontenc}
\usepackage{mathptmx}

\begin{document}

\preprint{AIP/123-QED}

\title{Effect of magnetic field on the lateral interaction of plasma plumes}

\author{Alamgir Mondal}
\email{alamgirphy@gmail.com}

\author{R. K. Singh}%

\author{Vishnu Chaudhari}

\author{H. C. Joshi}
\email{hem@ipr.res.in}
\affiliation{Institute for Plasma Research, Bhat, Gandhinagar, Gujarat, India - 382 428}

\date{\today}

\begin{abstract}
Lateral interaction between two geometrically modified plasma plumes in the presence of transverse magnetic field has been investigated. Characteristic behaviour of both seed plumes and interaction region in presence of field is compared with those for field free case. Contrary to the field free case, no sharp interaction zone is observed, rather large enhancement in emission intensities in both seed as well as interaction regions is observed in case of magnetic field. The observed results are explained on the basis of atomic analysis of the spectral lines from the interaction region of the interacting plumes. The physical processes responsible for higher electron temperature and increased ionic line emission from singly as well as doubly ionized aluminium are briefly discussed.
\end{abstract}

\maketitle

\section{\label{sec:level1}Introduction}
Collision of plasma plumes is an important phenomenon in many laboratory plasmas and has applications in plasma confinement, inertial confinement fusion (ICF), laboratory studies of plasma of astrophysical importance, generation of nano-particles and ion-sources etc.\cite{iftikhar2017partbeams, li2009optical, gekelman2003spacephy, ripin1987PRL, mostovych1989PRL} Expansion of laser produced plasma in the presence of an external magnetic field has been studied experimentally under different conditions and has been illustrated in recent works.\cite{dirnberger1994observation, behera2015confinement, singh2017mag, dawood2018magnetic, iftikhar2017partbeams, li2009optical, mostovych1989PRL} Harilal et. al. observed changes in plume shape and enhanced velocity and ionic emission in the presence of the magnetic field.\cite{harilal2007PRB} Behera et. al. observed oscillation in plasma plume.\cite{behera2015confinement}
Colliding plasma with different targets, laser parameters, ambient and ablation geometries has been reported by several authors.\cite{Ade2005early, SSharilal2001, gupta2013, ecamps2002, sanchez2006plume, Phough2010, XingwenLi2016, dardis2010stagnation, sangines2007time, kfalshabul, Bhupesh2014LBO, Bhupesh2015shock, Bhupesh2016ambienteffect} Kumar et. al. reported colliding plasma of thin film target and observed that neutral particles dominate in the interaction region.\cite{Bhupesh2013} In our earlier work in aluminium colliding plasma we had found a clear distinct interaction zone and neutral emission is significantly enhanced at later times due to increase in three body recombination.\cite{alampop2019}

The characteristics plasma parameters, geometrical shape and dynamics of the plasma plume are significantly modified in the presence of magnetic field. Therefore, it is quite interesting to see the plasma-plasma interaction in the presence of external magnetic field from the view point of its ramification in the plasmas of astrophysical importance to the fundamental and technological aspects. To the the best of our knowledge, study regarding the colliding plasma in the presence of an external magnetic field has not been attempted so far. In view of the above, we studied the behaviour of colliding plasma phenomenon in the presence of an external magnetic field by using fast imaging and optical emission spectroscopy(OES).


\begin{figure}
\begin{center}
\includegraphics[width = 7 cm]{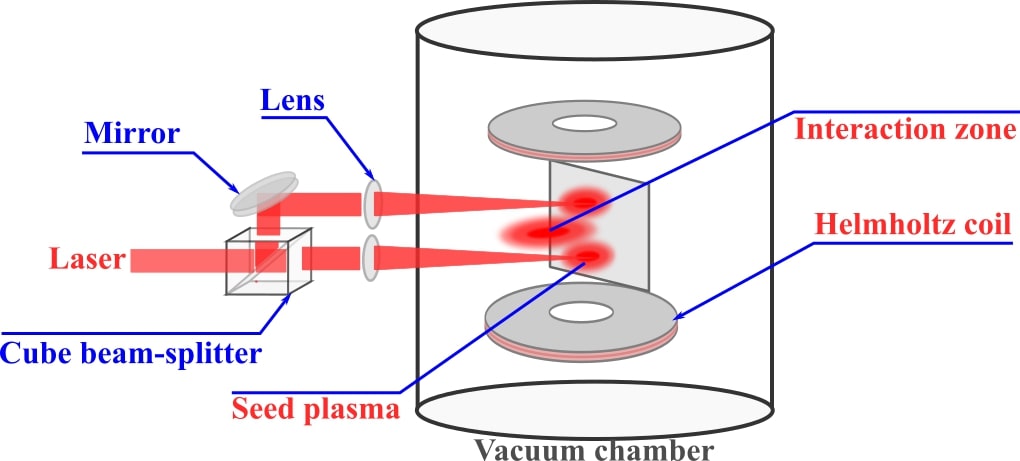}
\caption{\label{ch6:system} Schematic diagram of the colliding plasma in presence of magnetic field.}
\end{center}
\end{figure}

\begin{figure}
\begin{center}
\includegraphics[width = 8.5 cm]{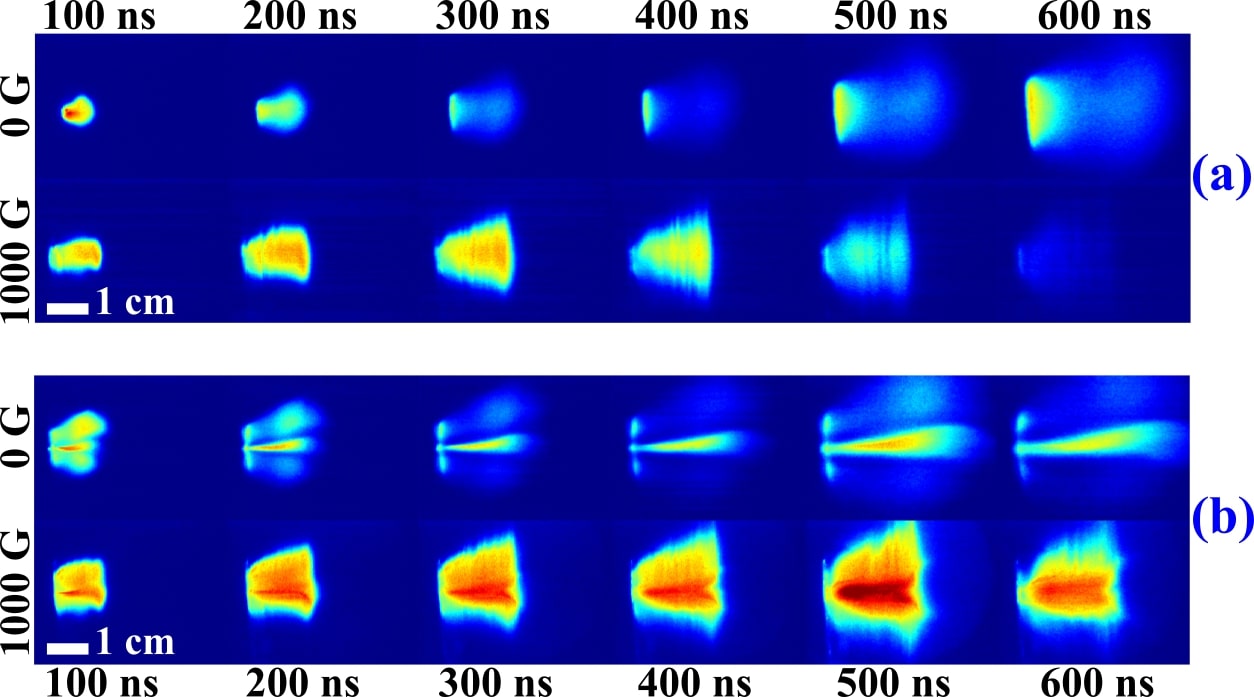}
\caption{\label{fig2:com} Temporal evolution of (a) single plasma (b) colliding plasma with and without presence of external magnetic field and 4 mm beam separation.}
\end{center}
\end{figure}

\begin{figure*}
\begin{center}
\includegraphics[width = 17 cm]{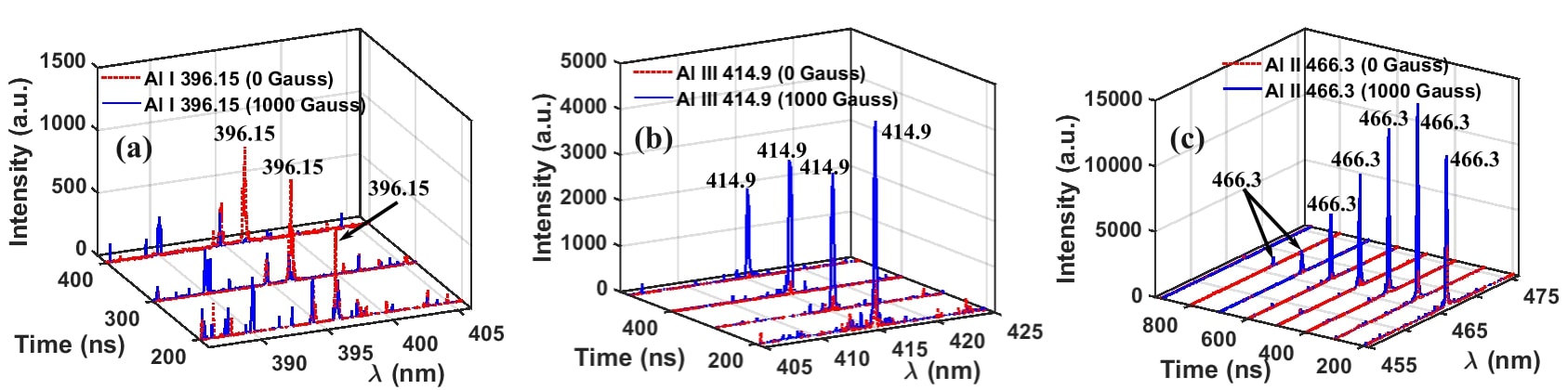}
\caption{\label{fig:Al_line}Temporal evolution of (a) Al I 396.15 nm, (b) Al III 414.9 nm and (c) Al II 466.3 nm lines with B = 0 and 1000 Gauss from interaction zone.}
\end{center}
\end{figure*}

\section{Experimental Setup}
A schematic diagram of the experimental setup is shown in Fig. \ref{ch6:system}. A specially designed Helmholtz coil which can produce a transverse magnetic field ranging from 0 to 6000 Gauss is placed inside a vacuum chamber. The coil has a flat-top magnetic profile (4.5 cm approximately at the center) which can be operated from outside without disturbing vacuum inside the chamber. Spatial flat top of the magnetic profile is always bigger than the maximum plume dimension in our experiment. Experiment is done in vacuum i.e. $5\times10^{-7}$ mbar. Detailed experimental discussions on experimental set-up are given elsewhere.\cite{alampop2019} Briefly, a 200 mJ laser beam of an Nd:YAG laser ($\lambda$ = 1064 nm, pulse width $\sim$8 ns, full-width at half-maximum) has been split into two beams of 100 mJ each. These two beams are focused by a plano-convex lens (35 cm focal length) on a clean aluminium target surface (99.9\% purity). The target dimensions are 6$\times$3$\times$1 cm$^3$ and is placed at the central region of the coil by using a vacuum compatible feed-through. Two laser shots are used to clean every new target position before recording any data. The target positions are changed by 2 mm to a fresh surface after each consecutive 5 shots by using the linear scale on the feed-through. The experiment is done in single shot mode. In case of images, we report the best repeatable image out of five shots. However, spectroscopic data and estimated plasma parameters are the average of five shots. The spot size and separation between the two beams are set as 1 mm and 4 mm, respectively. An ICCD (4 Picos, Stanford Computer Optics Inc.) camera with a time resolution of $\sim$200 ps is used to record plume images. A 0.5 m spectrometer (Acton Advanced SP 2500A) having overall resolution of 0.08 nm and coupled with an ICCD camera is used to collect the spatially resolved emission from the plasma plume at 3 mm away from the target surface by using a double lens telescopic arrangement. Magnetic field, ICCD camera, spectrograph and laser are synchronized with the combination of function generator and delay generator with jitter $\sim$ 1 ns.

\section{Results and discussion}
\subsection{Fast Imaging}
Figure \ref{fig2:com}(a) shows temporal evolution of the laser produced single plasma plume in field free case and in the presence of magnetic field. All the images in this article are spectrally integrated images in wavelength range of 350 to 700 nm and normalized to maximum intensity. The expansion of plasma without external field is free, adiabatic and its luminosity is beyond the detection limit at t>1000 ns. However, expansion dynamics and characteristics of the plasma changes with the introduction of transverse magnetic field. The major differences observed from the images are as follows. Due to higher initial kinetic energy of the plasma plume, its expand freely up to 300 ns. This phenomenon has been attributed to plasma oscillations due to diamagnetic effect\cite{bhadra1968oscillation, behera2015confinement}. After 300 ns, free expansion of the plasma plume in axial direction appears to be slowed down by the resistive force induced by external magnetic field. On the other hand, plasma plume does not experience any resistive force along the field lines and therefore plume expands freely along the magnetic poles. Well resolved striations along the field lines are observed in presence of field, which is more pronounced for higher fields and later stages of the plasma. Striation phenomenon has been studied by many authors in earlier works where it is mainly attributed as instability induced by shear velocity.\cite{behera2017twod, peyser1992} Another important feature observed in the presence of magnetic field is increase in emission intensity of the plasma and also the luminosity of the plasma plume persists up to a few microseconds. This is because electron temperature and density is significantly increased in the presence of field which will be discussed in latter section.

It can be seen from the above discussion that in the presence of external magnetic field, characteristics features, especially the geometrical shape of the plumes are significantly modified. Hence, it is interesting to carry out the comparative analysis of lateral interaction of the plumes in the absence and presence of magnetic field. Figure \ref{fig2:com}(b) represents time resolved images of lateral interactions of two spatially separated plasma plumes in the absence and presence of 1000 Gauss magnetic field, respectively, for time delays 100 to 600 ns in vacuum. In the absence of magnetic field, a well formed interaction zone is observed at the centre of interacting plumes which moves with higher velocity in comparison to the seed plumes. Interestingly the shape, size, geometry of the colliding plasma and the subsequent interaction zone exhibit drastic changes with introduction of the magnetic field. It can be seen from this figure that no clearly separated interaction zone is present in the presence of the field in contrast to field free case. It appears that the transversely elongated plumes in presence of magnetic field overlap each other. These changes can be understood as follows. Collisionality parameter i.e. $\zeta = D/\lambda_{ii}$ defines the nature of induced interaction zone in a colliding plasma where D is the separation between two laser beams and $\lambda_{ii}$ is ion-ion mean free path. Ion-ion mean free path is defined by the below equation.\cite{chenais1997kinetic}

\begin{equation}
    \lambda_{ii} = \dfrac{m_i^2 v_{12}^2}{4 \pi e^4 Z^4 n_i ln\Lambda_{12}}
\label{meanpath}
\end{equation}


\noindent Here, $m_i = 4.48 \times 10^{-23}$ g, $v_{12} = 2.3 \times 10^6 \quad cm/s$, Z = 2, $n_i = 1 \times 10^{16} \quad cm^{-3}$ and $\Lambda_{12} = 10$ are the ion mass, relative velocity of two plumes, ionization state, plasma density and Coulomb logarithm of the plasma. The estimated collisionality parameters are $\zeta$ = 5 and 10 for B = 0 and 1000 Gauss, respectively. These values predict soft stagnation for both the cases. However, in the presence of magnetic field, the ion-ion mean free path is likely to be modified because of ion gyration and hence can affect the estimated parameters. In this scenario the collisionality parameter may not represent the true picture of interaction region in the presence of the field. Further, this can be understood by the Larmor radii. Larmor radii are calculated from $m v_\perp / qB$, where m, $v_\perp$, q and B are mass of the charged particle, velocity perpendicular to the field, charge and magnetic field, respectively. Estimated values for electron, Al II and Al III Larmor radii are 3 $\mu$m, 9.9 cm and 4.4 cm, respectively. The Larmor radii for Al III ion is comparable to the plume dimension at later time delay and it is observed in the images as confinement of the plasma plume. However, we would like to mention that estimates of Larmor radii for Al III and Al II are done with coarse assumption that the velocity of both the species may be same which has been estimated from the temporally resolved images.

In the presence of the magnetic field, the value of $\beta$ plays an important role in governing the expansion dynamics of the plume. Thermal beta is expressed by the ratio of thermal pressure and magnetic pressure i.e. $\beta_t = \dfrac{n_e T_e}{B^2 / 2\mu_o}$, where all notations have their standard meanings. The expansion of the plasma plume transverse to the magnetic field stops, when thermal beta is equal to one i.e. thermal pressure of plasma is equal to magnetic pressure.  For the present experimental conditions, the estimated value of thermal beta is one at approximately 500 ns delay. However, plasma plume appears not to stop, rather it is slowed down as can be seen from Fig.\ref{fig2:com}(a) and \ref{fig2:com}(b). This is because in laser produced plasma, thermal energy is converted into directed energy and hence directed beta ($\beta_d$) becomes important. Directed beta is defined by the ratio of kinetic pressure to magnetic pressure i.e. $\beta_d = \dfrac{m n_e v^2 / 2}{B^2 / 2 \mu_0}$. The plasma expansion beyond the region $\beta_t$ $\approx$ 1, can be attributed to the directed beta of the plasma which is always greater than unity for the considered time delay.

\begin{figure}
\begin{center}
\includegraphics[width= 8.5 cm]{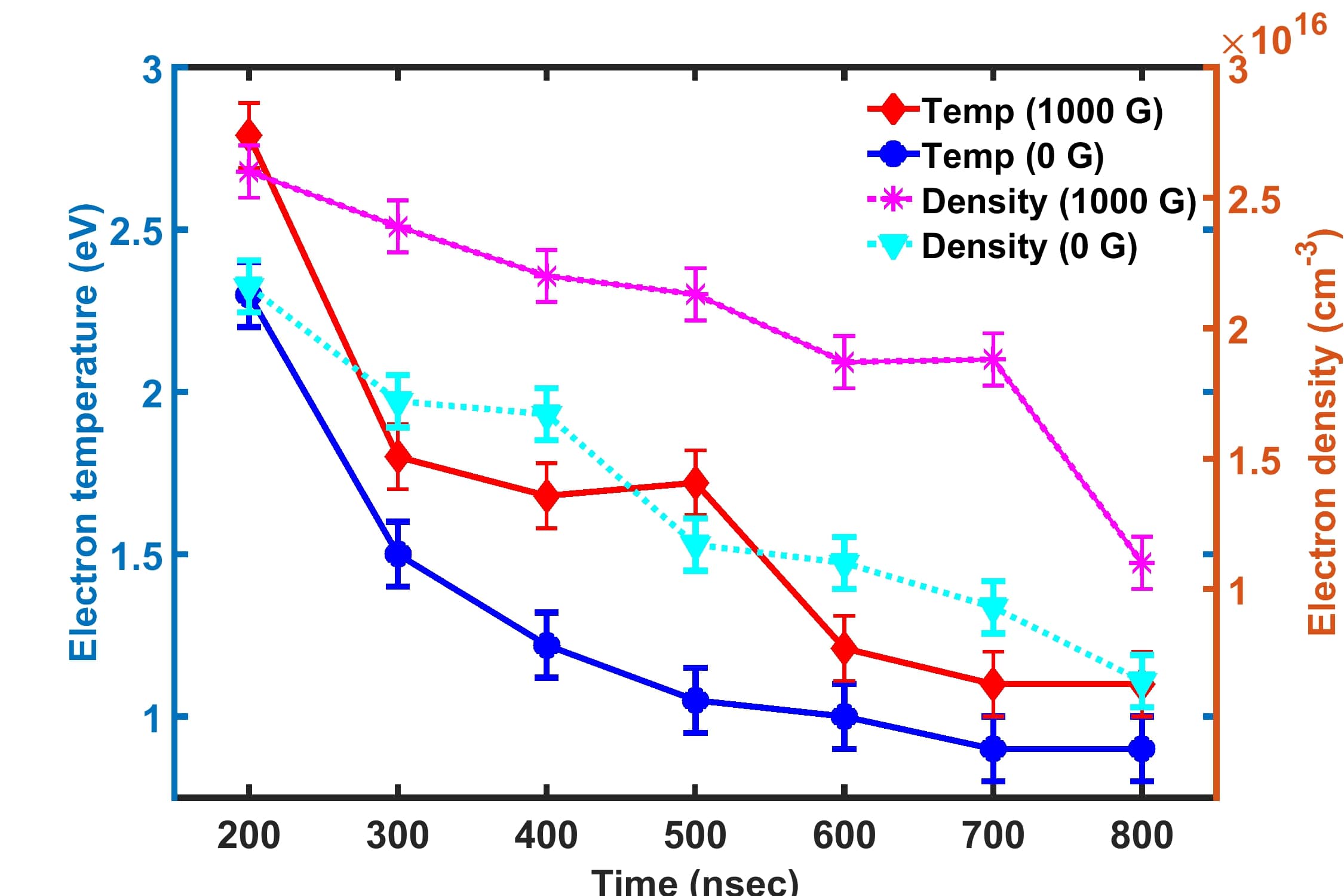}
\caption{\label{fig:temp_den} Temporal evolution of electron temperature and density of colliding plasma with field B = 1000 Gauss and without field.}
\end{center}
\end{figure}

\begin{figure}
\begin{center}
\includegraphics[width= 8 cm]{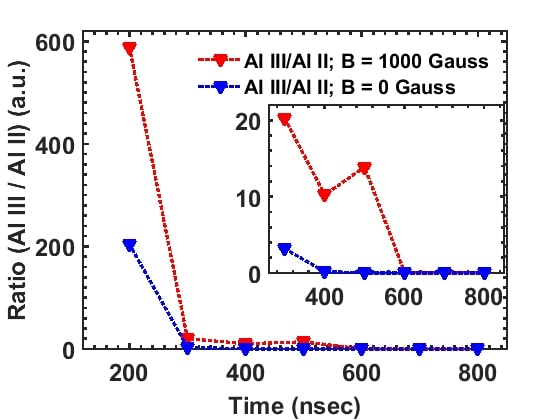}
\caption{\label{fig:ratio} Temporal evolution of Al III to Al II ratio calculated from Saha equation with B = 1000 Gauss and field free case. The inset figure shows the variation of the ratio between 300-800 nsec which is not clear from the larger figure.}
\end{center}
\end{figure}

\begin{figure*}
\begin{center}
\includegraphics[width= 15 cm]{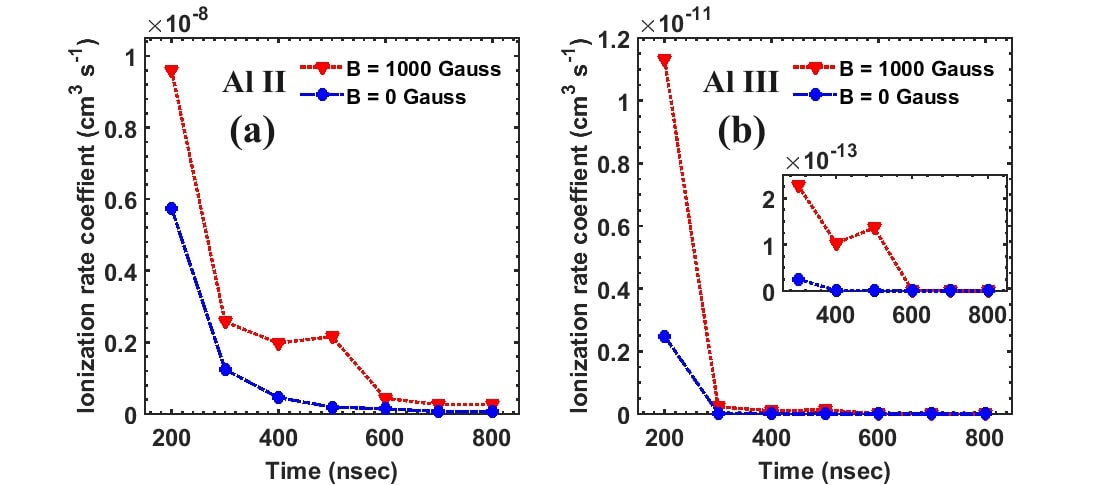}
\caption{\label{fig:ionrate} Temporal evolution of ionization rate coefficient for (a) Al II and (b) Al III.}
\end{center}
\end{figure*}

Another important parameter is bubble radius which is described by $R_b = [(3 \mu_o E_{lpp})/(2 \pi B^2)]^{1/3}$ for a spherical plasma plume expanding in magnetic field. Where, $\mu_o$ = $4 \pi \times 10^7$ H/m, E$_{lpp}$ is laser energy and B is the external magnetic field. The bubble radius, estimated from this equation is 1.77 cm for B = 1000 Gauss, which is comparable to our plume dimension of $\sim$2 cm at 300 ns where the maximum resistive force is observed with negligible axial velocity as observed in Fig. \ref{fig2:com}(a). The small difference observed in this case can be attributed to the assumption of spherical plasma plume, instead of ellipsoidal shape of the plasma plume.

Further, as in the case of single plasma plume, in colliding plasma also the expansion of the plasma plume in axial direction slows down in presence of field. Visual examination of Fig. \ref{fig2:com}(b) shows that luminosity of the plasma plume increases in the presence of field as compared to the field free case. This can be attributed to increased electron temperature and hence increase in ionic emission as will be discussed in the next section. These images clearly demonstrate that dynamics and shape of the colliding plumes, overall luminosity and formation of interaction zone is highly dependent on the presence of external magnetic field.

\subsection{Optical Emission Spectroscopy (OES)}
Optical emission spectroscopy is used to investigate plasma electron temperature, density and variation in intensities of lines from various charge states. Figure \ref{fig:Al_line} shows temporal changes in the intensity of characteristic Al I 396.15, Al II 466.3 and Al III 414.9 nm lines with B = 0 and 1000 Gauss. Al I line shows increase in intensity at longer times for field free case. Interestingly its intensity is considerably diminished in the presence of the field and almost unobservable at longer times which is in sharp contrast to the field free case. On the other hand, for ionic lines enhancement in intensity is observed. In contrast to the monotonic decrease in intensity in field free case \citep{alampop2019}, intensity of Al II increases up to certain value at 300 ns delay time, as seen for Al II 466.3 nm, and then starts decreasing with further time delays. The enhancement in the intensity of Al II ions is probably because of the increased temperature due to Joule heating in external field.\cite{harilal2004PRE} This is also supported by increase in the ionization rate coefficients with the introduction of magnetic field, which will be discussed latter. In the  forgoing discussion, it is shown that the electron density is also increased after the introduction of magnetic field, which is in line with our observation regarding increased ionization.

As mentioned earlier, it is interesting to note that the intensity of Al II and Al III lines increase whereas the intensity of Al I line decreases in the presence of field. To substantiate this we have estimated magnetic diffusion time described by $t_d = \dfrac{4 \pi \sigma R_b^2}{c^2}$, where $\sigma$ is plasma conductivity which can be estimated from Spitzer formula.\cite{nrl} The estimated magnetic diffusion time for R$_b$ = 1.77 cm and 2 eV temperature is 537 ns for Z = 2. This qualitatively explains the ionic emission even at longer times. This can be attributed to the increase in plasma electron temperature and subsequent density because of Joule heating in the presence of magnetic field. Again we would like to mention that this is in sharp contrast with field free case where neutral emission increases at longer times which has been attributed to increased three body recombination.\cite{alampop2019}

Electron temperature has been estimated by using Boltzmann relation which holds under local thermodynamic equilibrium (LTE). In LTE, excited states are populated according to Boltzmann distribution.\cite{ding_wu} Hence line ratio of two lines of a particular charge state can give the electron temperature.\cite{doria2006plasma, mraju2014influence, shen2007} The Boltzmann relation defined by the eq. \ref{eq_temp} and it is used for Al II 466.3 nm and 559.3 nm lines to estimate electron temperature..

\begin{equation}
    \dfrac{I_{ij}}{I_{kl}} = \dfrac{\nu_{ij}A_{ij}g_{i}}{\nu_{kl}A_{kl}g_{k}} exp^{-(E_i - E_k)/(k_{\beta}T_e)}
\label{eq_temp}
\end{equation}

\noindent In this relation, I is the line intensity of the transition between two energy levels, $\nu$ is the frequency of the line, A is Einstein's coefficient, g and E is statistical weight and energy of the particular energy level respectively, $k_\beta$ is Boltzmann constant, $T_e$ is electron temperature, and the subscripts i, j, k and l denote different energy levels. Electron density is estimated by Stark width of Al II 466.3 nm line as described in ref.\cite{alampop2019} Temporal evolution of electron temperature and density of colliding plasma with and without field are shown in Fig. \ref{fig:temp_den}. Electron temperature decreases with time delay in the presence of field and field free case. However, throughout this temporal evolution, the electron temperature is always higher in presence of magnetic field as compared to field free case. Electron density is increased in the presence of magnetic field and appears to show similar trend with time as in case of electron temperature.

McWhirter criterion for the validity of LTE condition is given by,\cite{bekefi}

$$n_e \geq 1.4 \times 10^{14} T_e^{1/2} (\Delta E)^3 \quad cm^{-3}$$

\noindent In the present experiment this criterion holds which requires electron density ($n_e$) to be greater than or equal to $1.5\times 10^{15}$ $cm^{-3}$, as the observed electron densities are always higher almost by an order (Fig. \ref{fig:temp_den}).

Measurement of line spectra of different charge states as shown in Fig. \ref{fig:Al_line} can not give the true picture of their abundance in the absence and presence of magnetic field. Therefore, in order to present the quantitative picture of different charge states in both the cases, we have estimated the ratio of Al III and Al II. Ratio of Al III and Al II ions has been calculated by using Saha relation described by eqn. \ref{eq:saha} and it is shown in Fig. \ref{fig:ratio} for both the cases i.e. with and without field.\cite{ffchen}

\begin{equation}
\label{eq:saha}
\dfrac{n_i}{n_n} \approx 2.4\times10^{21} \dfrac{T^{3/2}}{n_i} e^{U_i/{k_\beta T}}    
\end{equation}

\noindent Here, $n_i$, $n_n$, T, $U_i$, $k_\beta$ represents ion density, neutral density, plasma temperature in Kelvin, ionization potential of atoms and Boltzmann constant, respectively. Figure \ref{fig:ratio} clearly demonstrates the substantial increase in Al III ions with introduction of magnetic field.

To support the above observations, ionization rate coefficients for both Al II and Al III with and without external field has been estimated from the eqn. \ref{ratecof}. \cite{vriens1980rateoeff}

\begin{equation}
\label{ratecof}
\kappa_{pi} = \frac{9.56\times 10^{-6} (kT_e)^{-1.5} exp(-\epsilon_{pi})}{\epsilon_{pi}^{2.33} + 4.38\epsilon_{pi} + 1.32\epsilon_{pi}} \quad cm^3 \quad s^{-1}
\end{equation}

\noindent Here, $\epsilon_{pi} = E_{pi}/kT_e$ and $T_e$, $E_{pi}$, p, i represents electron temperature, ionization potential, principal quantum numbers of initial and ground state, respectively. Figure \ref{fig:ionrate} shows the temporal variation of ionization rate coefficients of colliding plasma at B = 0 and 1000 Gauss in vacuum. This figure shows that initially ionization rate coefficient decreases fast for both Al II and Al III lines with time. After that, it decreases at slower rate with further delay time. However, the increase in this rate is clearly visible with the introduction of magnetic field which describes fairly the spectral behaviour for neutral and ionic lines (Fig. \ref{fig:Al_line}) i.e. increased emission of Al II and significant intensity of Al III line even at later times as mentioned earlier.

Briefly two main striking observations are noticed in the interaction zone with the introduction  of magnetic field. First, no sharp interaction zone is observed which is expected from the estimation of collisionality parameter, instead a blurred overlapped region appears. This can be understood because of the gyration of ions in the presence of magnetic field. Second observation is increased emission from higher ionic states at longer times in contrast to increased neutral emission in field free case. This fact can be anticipated due to Joule heating effect in the presence of the magnetic field. 


\section{Conclusion}
It can be mentioned that the present study is an initiative in lateral interaction of laser produced plasmas in the presence of an external magnetic field. It can be mentioned that in the present work we report the spectral behavior as well as evolution of plume images at distance of 3 mm from the target. Interestingly contrary to field free case no sharp interaction region is noticed in the presence of magnetic field. Further, an increase of ionic emission, especially Al III emission has been observed in the presence of the magnetic field which is in sharp contrast to field free case where neutral emission dominates at longer times. This has been attributed to increased electron temperature and subsequent increase in ionization of plume species. Also the observed results are validated quantitatively by estimating the abundance of charge states and ionization rates for both the cases. We believe that the present work will be interesting from the view point of manipulating  colliding plasma properties with the introduction of magnetic field. Further study regarding the effect of spatial position, magnetic field and laser fluence is in progress which may bring more interesting aspects of the phenomena. Nevertheless two striking observations namely the absence of sharp interaction zone and appearance of higher charge states even at longer times are the important findings of the present work.

\section{data availability statement}
\noindent The data that support the findings of this study are available from the corresponding author upon reasonable request.
\section{Reference}
\nocite{*}

\providecommand{\noopsort}[1]{}\providecommand{\singleletter}[1]{#1}%

\end{document}